**Embedded EthiCS:**
**Integrating Ethics Broadly Across Computer Science Education**
Barbara J. Grosz[1], David Gray Grant[2], Kate Vredenburgh[2], Jeff Behrends[2], Lily Hu[3], Alison Simmons[2], and Jim Waldo[1]
[1]SEAS/Computer Science, [2]Department of Philosophy, and [3]SEAS/Applied Mathematics
Harvard University

**Abstract**

Computing technologies have become pervasive in daily life, sometimes bringing unintended but harmful consequences. For students to learn to think not only about what technology they *could* create, but also about what technology they *should* create, computer science curricula must expand to include ethical reasoning about the societal value and impact of these technologies. This paper presents *Embedded EthiCS,* a novel approach to integrating ethics into computer science education that incorporates ethical reasoning throughout courses in the standard computer science curriculum. It thus changes existing courses rather than requiring wholly new courses. The paper describes a pilot Embedded EthiCS program that embeds philosophers teaching ethical reasoning directly into computer science courses. It discusses lessons learned and challenges to implementing such a program across different types of academic institutions.

**Introduction**

Computing technologies are deeply intermeshed with the activities of daily life, playing an ever more central role in how we work, learn, communicate, socialize, and even participate in government. The ways in which these technologies are designed can have profound social implications. Despite the many ways they have improved life, computing technologies cannot be regarded as unambiguously beneficial or even value-neutral. Recent experience shows they can lead to unintended but harmful consequences. Some technologies are thought to threaten democracy through the spread of propaganda on online social networks, or to threaten privacy through the aggregation of data sets that include increasingly personal information, or to threaten just processes and outcomes when machine learning is used in high-stakes decision-making processes such as in loan applications, employment procedures, or parole hearings. It has become clear that it does not suffice to think about a technology's ethical consequences *after* it has produced negative social impacts, as has happened, for example, with facial recognition software that discriminates against people of color and with self-driving cars that are unable to cope with pedestrians who jaywalk. Developers of new technologies should aim to identify potential harmful consequences early in the design process and take steps to eliminate or mitigate them. This task is not easy. Designers will often have to negotiate among competing values – for instance, between efficiency and accessibility for a diverse user population, or between maximizing benefits and doing no harm. There is no simple recipe for identifying and solving ethical problems.

Computer science education can help meet these challenges by making ethical reasoning about the social impact of computing technologies a central element in the curriculum. Students could learn to think not only about what technology they *could* create, but also what technology they *should* create. Learning to reason this way requires different kinds of courses from those currently standard in computer science curricula. A range of university courses on topics in areas of computing, ethics, and public policy are emerging to meet this need. Some cover computer science broadly, while others focus on specific problems like privacy and security or on particular fields like artificial intelligence and data science. By and large, however, these courses are stand-alone courses in the computer science curriculum.

This paper presents an alternative and more integrative approach to incorporating ethical reasoning into computer science education, which we have dubbed "Embedded EthiCS". In contrast to one-shot, stand-alone computer-ethics or computer-and-society courses, Embedded EthiCS aims to make ethical reasoning an integral component of existing courses in the standard computer science curriculum.  It modifies existing courses rather than requiring wholly new courses. Students learn ways to identify ethical implications of technology and to reason clearly about them *while* they are learning ways to develop and implement algorithms, design interactive systems, and code, and not as a separate endeavor. Embedded EthiCS thus addresses issues with stand-alone courses raised in several studies of ethics in science and engineering education (Cech 2014; Hollander and Arenberg 2009). To provide expertise for teaching ethics similarly deep to that for the course technical content, this approach embeds graduate student and postdoctoral fellow philosophers directly into the teaching of computer science courses.

In the following sections, we elaborate on the rationale behind this approach; describe its development at Harvard, giving examples from a variety of participating courses; discuss lessons we have learned; and, consider challenges – intellectual, administrative, and institutional – to implementing such a program in academic institutions of different kinds. We conclude by calling for the computer science community to join together to build open repositories of resources to facilitate wider adoption of the approach.

**Why Embed Ethics and Philosophers in the Teaching of Computer Science?**

Embedded EthiCS was created in response to student demand for two elective courses in computer science at Harvard that considered ethical concerns in concert with computer science methods: "Privacy and Technology" and "Intelligent Systems: Design and Ethical Challenges." (For a brief description of these courses, see Appendix A.) In teaching these courses, we repeatedly noticed how easy it was for computer scientists to forget about ethical concerns when focused on technical systems issues. Even students earnestly committed to learning and using ethical reasoning in their work quickly lost sight of these considerations when engaged with a technical design task. At the same time, we recognized that most computer science courses contain material for which an ethical challenge might arise. We thus designed Embedded EthiCS to *habituate* students to thinking ethically, in contrast to approaches that interrupt students' technically-focused courses of study in computer science with a one-off course on ethics.

The Embedded EthiCS approach adds ethics mini-modules to computer science courses in the core computer science curriculum. By embedding ethics broadly across the computer science curriculum, this approach meets three goals for computer science students: it shows them the extent to which ethical issues permeate almost all areas of computer science; it familiarizes them with a variety of concrete ethical issues and problems that arise across the field; and, it provides them repeated experiences of reasoning through those issues and communicating their positions effectively.

While no single course with an Embedded EthiCS module will by itself produce ethically-minded technology designers, we expect that incorporating modules throughout the curriculum will have a compounding effect—one that continually reinforces the importance of ethical reasoning to all aspects of computer science and technology design. Adding ethics modules across curricular offerings also exposes students to ethical content in a great breadth of computational contexts. This approach thus promises to convey the message that ethical reasoning is an expected part of computer scientists' work.

Embedded EthiCS is inherently interdisciplinary. Knowing what *can* be done with technology falls within the purview of computer science. Understanding, evaluating, and successfully defending arguments about what *should* be done falls within the purview of the normative disciplines, most notably ethics, a subfield of philosophy. For students to succeed at learning not only *how* to build innovative computing systems, but also the ways to determine what systems they *should* build, it is imperative that these two disciplines work together. To this end, Harvard Computer Science faculty have been partnering with the Harvard Philosophy Department to develop the Embedded EthiCS curriculum. To enable us to scale the program up to the full computer science curriculum, Computer Science faculty and teaching assistants collaborate with advanced Ph.D. students and postdoctoral fellows in Philosophy to develop Embedded EthiCS modules for each course. This approach also opens up exciting new areas of research for the philosophers who teach the mini-modules and broadens their teaching repertoire considerably.

**How Does Embedded EthiCS Work?**

Each Embedded EthiCS course has an Embedded EthiCS teaching assistant who is an advanced Ph.D. student or postdoctoral fellow in Philosophy with a strong background in ethics and considerable teaching experience. The teaching assistant works with the faculty course head to design an ethics mini-module in which students acquire practical competence in thinking through particular ethical challenges. Together, the faculty member and teaching assistant identify a topic covered in the course that raises an interesting ethical issue, develop case studies to be used in one or two class sessions, and design an assignment on the material. The Embedded EthiCS fellow is responsible for preparing and leading the class sessions, as well as for designing the assignment and a way to assess it. Depending on class size, the grading itself may be done by the Embedded EthiCS teaching assistant, by regular course teaching assistants, or through peer grading.

The mini-modules are designed to give students three core ethical reasoning skills: the ability to identify and anticipate ethical problems in the development and use of computing technologies; the ability to reason, both alone and in collaboration with others, about those problems and potential solutions to them, using concepts and principles from moral philosophy; and the ability to communicate effectively their understanding and approaches to addressing those problems. The modules emphasize "active learning" activities and assignments that teach students to apply the philosophical ideas they have learned to concrete, real-world ethical problems as recommended by recent studies of ethics education (Cech 2014; Hollander and Arenberg 2009). They are designed to help students exercise their newly acquired ethical reasoning skills in context. The assignments encourage practice in articulating a well-formed analysis of the ethical problem(s) in context, developing a solution based on that analysis, and providing clear and persuasive reasons in defense of that solution, deepening students competence in this type of analysis.

**Embedded EthiCS Pilot**

We piloted the Embedded EthiCS program over three semesters (Spring 2017, Fall 2017 and Spring 2018), with fourteen separate courses participating (three of them twice). Figure 1 lists the courses, grouping them by Computer Science area and indicating the ethical problems discussed. To illustrate the content and design of mini-modules, we describe the four courses in bold in more detail below.

| AREA | TITLE | CHALLENGES | COURSE # |
|---|---|---|---|
| Introductory Courses | Great Ideas in Computer Science | The Ethics of Electronic Privacy | CS 1 |
| | Introduction to Computer Science II | Morally Responsible Software Engineering | CS 51 |
| | Advanced Topics in Data Science | Moral Considerations for Data Science Decisions | CS 109B |
| Theory | Fairness, Privacy, and Validity in Data Analysis | Diversity and Equality of Opportunity in Automated Hiring Systems | CS 126 |
| CS and Economics | Networks | Facebook, Fake News, and the Ethics of Censorship | CS 134 |
| | Economics and Computing | Matching Mechanisms and Fairness | CS 136 |
| | Topics at the Interface of Economics and Computing | Interpretability and Fairness | CS 236R |
| Programming Languages and Computer Systems | Programming Languages | Verifiably Ethical Software Systems | CS 152 |
| | Data Systems | Data and Privacy | CS 165 |
| | **Big Data Systems** | Privacy and Statistical Inference from Data | CS 265 |
| Human-Computer Interaction | **Design of Useful and Usable Interactive Systems** | Inclusive Design and Equality of Opportunity | CS 179 |
| Artificial Intelligence | **Machine Learning** | Machine Learning and Discrimination | CS 181 |
| | Introduction to AI | Machines and Moral Decision-Making | CS 182 |
| | Autonomous Robot Systems | Robots and Work | CS 189 |

**Figure 1: Embedded EthiCS courses 2017-2018. CS236R and CS265 are graduate courses; other courses are primarily for undergraduates. Boldface courses are described below.**

1. **Networks: Facebook, Fake News, and the Ethics of Censorship** (primarily for undergraduates)

This course focuses on the use of network modeling tools to study complex empirical phenomena of current online networks, including ways ideas and influence spread, and the contagion of economic behavors. The Embedded EthiCS mini-module considered the question of censorship of so-called "fake news" by social media companies. Its goal was to introduce students to engage in different forms of ethical reasoning about questions of whether social media companies are morally obligated to suppress the spread of "fake news" on their platforms, and, if so, what kinds of content they should suppress and what strategies they should use to suppress it. The Embedded EthiCS teaching assistant first discussed three philosophical theories and issues with the students: the distinction between hard and soft censorship; a selection of J.S. Mill's arguments against censorship from *On Liberty* (Mill 1859); and, an argument, reconstructed from a *New York Times* editorial, that Facebook is obligated to suppress fake news because it interferes with democratic governance (New York Times Editorial Board 2016). The module's assignment asked students to write short essays identifying a strategy for suppressing fake news and defending a position about whether Facebook was obligated to implement it.

2. **Big Data Systems: Privacy and Statistical Inference** (for graduate students and advanced undergraduates)

This discussion-based course investigates the design of data systems and algorithms that can "scale up," i.e., use a single machine to its full potential, and "scale out," i.e., use multiple machines (typically in the hundreds or thousands). The Embedded EthiCS mini-module considered how to understand and protect privacy in the age of big data, particularly in light of the powerful inference capabilities large data sets and contemporary analytical tools make possible, some of which seem to violate individual privacy. Its goals were to give students a methodology for diagnosing the importance of privacy in a domain; to help students understand why traditional privacy protections, such as consent notices and anonymization, are ineffective for some flows of information; and to have students brainstorm solutions to difficult cases of statistical inference from publicly available information. To prepare for the in-class discussion, students were assigned a set of detailed questions on readings that dealt with different definitions of privacy and types of privacy protections (Rumbold and Wilson 2018; Barocas and Nissenbaum 2014; Levy and Barocas 2018; Dwork 2006). In class, the Embedded EthiCS teaching assistant introduced an interest-based methodology for thinking about these issues (Scanlon 2009, Tasioulas 2007 inter alia). The methodology starts by identifying the serious, common interests that underlie rights protections. The in-class session focused on the ethical grey area of whether unforeseen inferences about an individual from her publicly available data constitute a violation of privacy. (See Rumbold and Wilson 2018 for discussion.) The class also discussed whether individuals did or did not waive their right to privacy in other grey areas, such as when employers monitor employees at work. For cases where privacy was violated, students brainstormed design solutions using the methodology.

3.     **Design of Useful and Usable Interactive Systems: Inclusive Design and Equality of Opportunity** (primarily for undergraduates)

The Embedded EthiCS mini-module for this human-computer interaction design course focused on the topic of inclusive design, viz., designing human-computer interaction systems to be both useful to and usable by individuals with disabilities of various kinds. Its goal was to lead students to think more clearly about whether, and to what extent, software developers are morally obligated to design for inclusion. The class began with a discussion of different meanings of "inclusive design." Students then considered whether software companies are morally obligated to design for inclusion because doing so would, at a reasonable cost, alleviate unjust cumulative disadvantages faced by people with disabilities. During this discussion, the Embedded EthiCS teaching assistant introduced three relevant philosophical ideas: the distinction between actions that are morally obligatory and morally supererogatory; John Rawls's principle of fair equality of opportunity (Rawls 1971); and, the medical, social, and interactive models of disability (Wasserman et. al. 2016). Theory was put into practice in a class activity: students engaged in a group-based ethics simulation in which they imagined that they were software developers deciding whether to redesign their company's website for inclusion even if they might incur a cost like doing the work on personal time. The module's assignment was incorporated into the final design project for the course. Students were asked to answer questions about whether they would be obligated to design their project for inclusion if they went on to develop it commercially.

4.     **Machine Learning and Discrimination** (primarily for undergraduates)

The Embedded EthiCS mini-module for this introductory machine learning course focused on machine learning and its potential for discrimination. Its goals were to introduce students to different theories of wrongful discrimination; to lead students to appreciate that designing ethical machine learning systems involves more than designing accurate machine learning algorithms; and, to introduce students to formalized fairness criteria and lead them to think about the implications of an "impossibility" result (Kleinberg, Mullainathan, Raghavan 2016). After giving a brief presentation on theories of wrongful discrimination (for which Hellman 2011, Chapter 1 provides an overview), the Embedded EthiCS teaching assistant presented a case study in which an employer's hiring practices generated outcomes that correlated with the race of job applicants (based on Barocas and Selbst 2016). The procedure was grounded in a sound business rationale and was also the product of historical injustice against certain groups. The students discussed whether the case was an instance of discrimination on two different types of theories of discrimination: anti-classification theories and anti-subordination theories (Barocas and Selbst 2016). The distinction between these two theories was then used to discuss conflicts between formal fairness criteria and the public discussion surrounding the use of COMPAS, a recidivism risk prediction tool, to inform judge's decisions in parole hearings. As an assignment, students were asked to design an algorithm for fair hiring practices that would reduce disparate impact while also producing socially good outcomes in the labor market, and to defend their design choices.

**Embedded EthiCS: Assessment of the Pilot-Program Class Modules**

Our experience with the pilot program has shown that it is not only possible to integrate the teaching of ethical reasoning with core computer science methods but also rewarding for students and faculty alike. Following each Embedded EthiCS class session, faculty informally provided feedback, and we asked students to complete a short survey. Faculty reported that mini-modules contributed to classes with only a modest burden on them, and that they learned from the modules.

Student surveys aimed to assess the effectiveness of each module and of the mini-module approach in general. Appendix B presents key survey results. Responses were overwhelmingly positive, supporting continuation of the initiative. The results show over 80% of students in all courses—and over 90% of students in five of the classes—agreed that these class sessions were interesting. In all but two classes, more than 80% of students reported that they would be interested in learning more about ethics in future computer science courses. Comments, which one quarter of the students provided, were overwhelmingly positive, with many expressing eagerness for more exposure to ethics content and more opportunities to develop skills in ethical reasoning and communication. Negative comments were largely specific to individual class content or presentation. Some students wanted more breadth or depth, others more background. One comment about overlap between two classes suggests the need to coordinate across classes.

**From the Pilot to a Sustainable Program**

For the first pilot of Embedded EthiCS in the spring semester of 2017, one Ph.D. student developed modules for four different classes: the introductory "great ideas" class, a theory of networks course, a data science course, and a human-computer interaction class. Based on the success of that effort, in AY 2017-18, we engaged two Ph.D. students, who added modules to an additional 10 courses in addition to continuing the modules in 3 of the original 4 courses.

In AY 2018-2019, we will be working toward developing a corps of graduate student and postdoctoral teaching assistants for the program. A postdoctoral fellow will lead weekly meetings of past, present and potential future teaching assistants and coordinate their development of mini-modules. In Fall 2018, we expect nine courses to include Embedded EthiCS modules, including three new courses, two in systems and one in theoretical computer science.

What have we learned? The key lessons concern student engagement, faculty roles, teaching assistant experience, and barriers to embedding ethics. A set of best practices is emerging.

For engaging students with ethical reasoning, we found that the Embedded EthiCS approach works best when the issues raised in the class session connect technical material to ethical issues already salient to students (e.g., privacy), the module employs short active learning participatory activities throughout the class session, and an assignment gives students practice with the ideas developed in the session.

We found four key roles for Computer Science faculty in ensuring the success of Embedded EthiCS modules: their participating fully in the design of the Embedded Ethics mini-module; their commitment to having an assignment (either separate or part of a larger problem set) that contributes to the course grade in some way, however minor; their being present and personally involved in the Embedded EthiCS module class session(s); and their raising ethical issues during other parts of the course either to preview the upcoming module or to refer back to the lesson. When the assignment contributes to the final course grade and when faculty are physically present in the Embedded EthiCS class session, students understand that the faculty value the place of ethical reasoning in the course and that the module is a core element of the course content rather than an optional supplement.

We have found that Ph.D. students and post-doctoral fellows who are teaching assistants for Embedded EthiCS can embed modules in three to four different courses per term, depending on how many modules are new and how much material is already available. Although the work is very different from the typical Philosophy teaching assistant's leading of discussion sections and grading essays, preparing and teaching three or four Embedded EthiCS modules is roughly the workload of teaching two discussion sections. Further, the teaching assistants who have participated to date have reported that they benefited enormously from the exposure to a breadth of computer science concepts and methods for which their philosophical expertise is relevant. We anticipate that this experience will also prove valuable on the job market and, for many, to their research.

Several of the barriers we encountered are common within cross-disciplinary ventures. First, we saw typical insecurities: philosophers who were concerned about their lack of technical expertise and computer scientists who were concerned about their lack of familiarity with ethics and reluctant to discuss ethical issues with students. Although we found that the technical barrier to productive ethical discussions of computer science methods and systems is much lower than many philosophers expect, philosophers without a background in computer science still need support, both financial and intellectual, to develop the requisite background knowledge. We also found many Computer Science faculty interested in attending a brief introductory computer ethics course focused on philosophical theories and methods relevant to computing technology challenges, a possibility that we are currently exploring.

Our experience suggests that the process of co-designing Embedded EthiCS modules helps mitigate insecurities, and the presence in the class of philosophers with the expertise to answer questions is essential for success. Strategies we have found helpful include selecting the topic for the ethics module before the semester begins, with input from both computer science faculty and philosophers. Doing so provides the philosopher time to develop the required technical knowledge in the relevant specific content area. Building a repository of past material for reuse or to serve as models for future modules has also proved useful.

Second, as is also typical in cross-disciplinary efforts, the different vocabularies and methodologies of computer science and philosophy can lead to confusion on the part of students and to interactions that amount to talking past one another. Students also can have trouble addressing problems posed in a computer science course that do not have a single correct answer. We found it important to involve computer science faculty and teaching assistants throughout the design and implementation of Embedded EthiCS modules to create more overlap between the ethics material and the computer science course vocabulary, and to anticipate the philosophical material in need of more explanation.

A third cross-field challenge is recruiting philosophers to develop and teach the Embedded EthiCS mini-modules. Hiring numerous Ph.D. candidates from a single philosophy graduate program to cover the full range of computer science courses is impractical. The Philosophy Department needs these graduate students to teach its courses, and the students themselves need experience teaching those courses. To address this challenge, we are including philosophy postdoctoral fellows in the teaching assistant cadre. For these postdoctoral fellows too, we expect the training and experience they gain will benefit their research and employment profile.

We also uncovered assessment and institutional challenges. The approach of integrating ethics pedagogy into core computer science courses reflects a hypothesis that recurring exposure to this type of reasoning across the curriculum will habituate students to thinking ethically when pursuing technical work. Post-module surveys provide insight into the effectiveness of particular modules, but we want to measure the approach's impact over the course of years, for instance as students complete their degrees and even later in their careers. We thus need to find ways to measure the long-term effectiveness of the Embedded EthiCS approach and compare it to other approaches.

The institutional challenges to mounting Embedded EthiCS derive from its cross-disciplinary nature. In particular, university support, both financial and administrative, is crucial. Funding is needed for teaching assistants and postdoctoral fellows, including senior level postdoctoral fellows able to train and support the efforts of those developing modules for courses. Administrative support is needed for recruiting faculty and courses in Computer Science, for recruiting teaching assistants and postdoctoral fellows in Philosophy, and for organizing and managing a repository of materials for the program, including mini-modules and evaluation materials. Several of these challenges are made more complex because they cross university divisions.

**Looking Forward**

Teaching computer scientists to identify and solve ethical problems starting from the design phase is as important as enabling them to develop algorithms and programs that work efficiently. The strategy of integrating the teaching of ethical reasoning skills with the teaching of computational techniques into existing computer science coursework not only provides students valuable experience identifying, confronting, and working through ethical questions, but also communicates the need to identify, confront, and address ethical questions throughout their work in computer science. It provides them with ethical reasoning skills to take into their computing and information technology work after they graduate, equipping them to produce socially and ethically responsible computer technology, and knowledge of ways to argue with those with and for whom they work for their ethically-motivated design choices. Computer scientists and technologists with these capabilities are important for the long-term well-being of our society.

We invite those at other institutions to join us by integrating ethics throughout their own computer science curriculum and to help us expand the open repositories of resources we are developing for ethics mini-modules, including in-class activities, assignments, and recommended readings. We also think it is important to share lessons learned, approaches to meeting the challenges of university support for these efforts, and ways to engage and train philosophers to participate in them.

**Appendix A**

The elective computer science courses "Privacy and Technology" and "Intelligent Systems: Design and Ethical Challenges" are broadly multidisciplinary and integrate significant amounts of ethical material with computer science material. They admit students with varying backgrounds in computer science (from sophomores who have taken only the introductory programming course to seniors majoring in computer science) and academic interests (from literature and policy to computer systems), who nonetheless share an interest in technology and its influence on individuals and society. The courses are small (24-36 students) to enable in-depth discussions and interaction. Student interest has been high and ever increasing; in recent years, 140-150 students have applied to each course, which is all the more remarkable as applications require substantive content.

In this appendix, we describe briefly these two courses, which integrate ethics and computer science material more fully, to illustrate a more extensive integration of ethics with computer science. Embedded EthiCS may be viewed to some extent as aiming at the same extensiveness in a distributed manner. Although the courses cover very different areas of computer science, they share many features in common. The teaching staffs include faculty and teaching assistants with expertise in computer science and in ethics. This enables them to engage responsibly with the full range of student questions on both aspects of the course. Class sessions and assignments integrate computer science technical content with ethical analysis, so that students may come to understand both kinds of concepts and their interrelationships.

**1. Privacy and Technology**

The course on "Privacy and Technology" examines technological advances that pose challenges to various intuitive notions of privacy and the laws and policies that are meant to protect privacy. The course aims to educate students who will go on to careers in law and policy about ways to think about the technology, and to teach those who will go on to careers in technology how to recognize potential privacy violations and ways those violations can be addressed through a variety of approaches, technological, legal, and regulatory.

Students examine particular technologies with the aim of understanding both what the technology is capable of doing and whether or not it poses a genuine threat to privacy. If it does, they consider whether the possible solutions are best achieved by a change in the technology or by new laws or policies. Assignments include writing of position, policy, or briefing papers and technology assessment exercises, both typically done by small groups. Students learn to explain technology to policy-centric audiences and to design technologies and policies that would govern them to minimize privacy invasion. Technology assessment assignments include geo-tagging all surveillance cameras on campus and analyzing a science-fiction movie deploying privacy-invasive technology to determine the time scale on which it might be realized and the advances needed to make it so.

Final projects in the course have included attempting to re-identify public data sets that are purportedly "de-identified", the creation of maps of drug-related photographs used in advertisements on the Silk Road dark-web site, legal analyses of privacy law regimes in various countries, and in-depth studies and analyses of the Indian Aadhar system.

**2. Intelligent Systems: Design and Ethical Challenges**

This course aims to combine a broad introduction to artificial intelligence including both its current and potential future uses, with strategies to address the design and ethical challenges it raises. Course readings, discussions, and assignments cover a range of AI methods that enable students to understand the ways AI technologies work and, by experimenting with various algorithmic tools and systems in common use, to identify their strengths and weaknesses. They combine theoretical and applied ethics content to provide students with a set of tools for developing their ethical awareness as well as a range of applied arguments with which to engage and hone their own argumentative skills.

We typically divide the course into four sections, each of which covers a particular area of AI, with readings, class discussions, and assignments in each section interweaving technical topics and ethical issues. For example, in covering planning and decision-theoretic reasoning, the autonomous agent decision-making module of the course includes readings and class sessions on Markov decision processes and reinforcement learning, their application to design challenges in autonomous vehicles (AVs), an introduction to virtue ethics, and applied ethical material about how AVs should behave in scenarios involving unavoidable collisions and how social policy interacts with that question. For assignments in this module, students empirically investigate reinforcement learning approaches that bear on technical AV design and undertake an argumentative analysis of an ethical challenge for autonomous agent decision-making in a medical, legal, or policing context.

Early in the semester, we introduce students to three broad approaches to ethical theorizing from the philosophical tradition. This theoretical material is supplemented with appropriate readings in applied ethics. Students learn that careful ethical thinking is not merely a matter of learning a series of theories and then becoming adept at applying them individually to particular cases. Rather, familiarity with ethical theories is useful for becoming attuned to particular features of cases that may be ethically significant and ways to approach thinking about them.

In their final projects, students work (typically in pairs, though for exceptional projects individually) on a project that takes as its *primary* focus either a technical or ethical problem but which includes attention to both, thus integrating the technical and normative thinking skills they have developed. For example, in one technically focused project, two students applied Naive Bayes classifiers to build a natural language processing system able to serve as a personalized assistive running coach. Their final project report described the system design and provided an ethical analysis of the key privacy considerations of personalized exercise data as well as potential health-related side effects of using such athletic training devices. In another ethically-focused project, a student undertook an extensive ethical analysis of news recommender systems and then proposed several design alternatives for addressing "fake news".

## Appendix B: Embedded EthiCS Pilot Evalution Summary

| STATEMENT | PERCENTAGE OF STUDENTS WHO AGREED WITH STATEMENT | | | | | | | | | |
|---|---|---|---|---|---|---|---|---|---|---|
| | CS 1 | CS 51 | CS 109B | CS 134 | | CS 136 | CS 152 | CS 165 | CS 179 | CS 182 |
| The ethics guest lecture was interesting. | 96% | 95% | 81% | 93% | 100% | 86% | 86% | 100% | 83% | 80% |
| The ethics guest lecture was relevant to me. | 91% | 86% | 90% | 89% | 100% | 86% | 78% | 100% | 89% | 80% |
| The ethics guest lecture helped me think more clearly about the moral issues we discussed. | 91% | 98% | 76% | 87% | 80% | 71% | 78% | 100% | 83% | 60% |
| The ethics guest lecture increased my interest in learning about the moral issues we discussed. | 83% | 90% | 86% | 84% | 87% | 86% | 81% | 100% | 72% | 80% |
| I would be interested in learning more about ethics in future computer science courses. | 83% | 83% | 90% | 85% | 73% | 86% | 76% | 100% | 74% | 100% |

**Figure 2:** Percentage of responding students in each course who agreed with each statement from the student evaluation survey. (Note that original responses were on a Likert scale from 1-7, with 7 = "strongly agree," 6 = "agree," 5 = "somewhat agree," 4 = "neither agree nor disagree," 3 = "somewhat disagree," 2 = "disagree," 1 = "strongly disagree.") CS134 was offered twice, and results from both surveys are provided in chronological order. CS 179 was also offered twice; we show the initial survey results; the subsequent survey had a higher percentages in all categories. Figure 1 may be consulted for course titles and the ethical challenges discussed in each course.